\documentclass[runningheads]{llncs}
\usepackage{amsmath}
\usepackage{booktabs}       
\usepackage{dsfont}
\usepackage{enumerate}
\usepackage{hyperref}       
\usepackage{makecell}
\usepackage{tabularx}

\newcolumntype{Y}{>{\centering\arraybackslash}X}
\makeatletter
\newcommand*\bigcdot{\mathpalette\bigcdot@{.5}}
\newcommand*\bigcdot@[2]{\mathbin{\vcenter{\hbox{\scalebox{#2}{$\m@th#1\bullet$}}}}}
\makeatother

\usepackage{graphicx}
\begin{document}
\title{The Impact of Incomplete Information on Network Formation with Heterogeneous Agents}
%
\titlerunning{Uncertainty in Social Networks}
%
\author{D. Kai Zhang\orcidID{0000-0001-5770-0903} \and
Alexander Carver\orcidID{0000-0003-3151-6563}}

%
\authorrunning{D.K. Zhang, A. Carver}

%
\institute{Department of Computing \\ Imperial College London \\
\email{\{dz819, alexander.carver15\}@imperial.ac.uk}}
\maketitle              
\begin{abstract}
    
We propose an agent-based network formation model under uncertainty with the objective of relaxing the common assumption of complete information, calling attention to the role beliefs may play in segregation. We demonstrate that our model is capable of generating a set of networks that encompasses those of a complete information model. Further, we show that by allowing agents to be biased toward each other based on observable attributes, our model is able to generate homophilous equilibria with preferences that are indifferent to these attributes. We accompany our theoretical results with a simulation-based investigation of the relationship between beliefs and segregation and show that biased beliefs are an important driver of segregation under incomplete information. 


\end{abstract}
\section{Introduction}



Frequently, members of the same group exhibit similar outward appearances -- a phenomenon referred to as homophily. This paper aims to provide an approach to studying this feature of social networks under uncertainty. We specifically build on an agent-based network formation model first proposed by Jackson and Wolinsky (1996) \cite{Jackson1996}. 

A number of extensions have been made to study segregation from a network formation perspective. These extensions have in common that agents belong to different types, where a type denotes shared attributes. Further, agents may exhibit heterogeneity in either benefits or costs when connecting to other agents. Specifically, forming a link with someone of the same type is usually assumed to either offer greater benefits or come at a lower cost than forming a link when types differ. By way of illustration, these models imply that a person finds it easier to form friendships with other people of the same ethnicity. These models show that segregation equilibria exist when agents exhibit even small preferences for homophily. 

\subsection{Our Contribution}
The key shortcoming in most of the existing literature is the strong assumption that agents have complete information about each other’s types or, equivalently, utilities depend only on observable attributes. This paper departs from this assumption by introducing a distinction between overt social groups, which subsume all publicly observable attributes, and covert types, denoting private attributes. Agents are assumed to prefer connecting to other agents of the same type which, however, is \textit{ex-ante} unobservable. Agents are thus indifferent over observable attributes. Instead, agents use a heuristic device and rely on social group memberships to gauge someone’s hidden type. This, in turn, allows for “errors” in people’s judgement of each other’s types.

We present a model designed to give us the ability to investigate if segregation between social groups can simply occur when people are sufficiently prejudiced, i.e., exhibit beliefs that are detached from the true distribution of covert types. We introduce the concept not just of hidden types, but also of agents' beliefs to intuit these types from observable attributes and form connections based on these beliefs.

Finally we implement various simulated experiments in order to provide comparisons to complete information networks and incomplete information networks with rational beliefs. We demonstrate that even mildly biased beliefs are an important driver of segregation under incomplete information.



\subsection{Related Literature}

Since Schelling's (1969, 1971) \cite{Schelling1969,Schelling1971} landmark segregation models, numerous studies of homophily in heterogeneous networks have been undertaken in recent years. Jackson and Xing (2014) \cite{JacksonXing2014} argue that homophily reflects how norms could help members of a group predict each other’s behaviour and thus facilitate cooperative play in coordination games. McPherson, Smith-Lovin and Cook (2001) \cite{McPherson2001} show that having similar attributes is a strong predictor that two individuals maintain social ties along a variety of dimensions. 

Aits, Carver and Turrini (2019) \cite{DBLP:conf/atal/AitsCT19} demonstrate how tolerance can result in complex intra-group segregation. De Mart{\'{i}} and Zenou (2017) \cite{DeMarti2017} determine that segregation can be the socially optimal outcome when agents are not sufficiently exposed to each other's groups as to reduce inter-group linking costs. Centola, Gonzalez-Avella, Eguiluz and San Miguel (2007) \cite{Centola2007} show how homophily can lead to the formation of distinct cultural groups that persist even when random shocks variously bring groups closer together. Within this divergence, group membership can have profound effects on the individuals. This has been examined, amongst others, in the context of migration and assimilation decisions \cite{Blumenstock2019,Chay2013,Munshi2016,Verdier2017}, occupation and employment \cite{Granovetter1983,Munshi2011,Zenou2015}, and even marriage \cite{Skopek2011}. 

However, to date, only a relatively small body of literature exists which has investigated the role of incomplete information in networks. Jackson and Yariv (2007) \cite{JacksonYariv2007} investigate incomplete information pertaining to the structure of the network. They present a model in which agents are uncertain over how well connected other agents are. Since an agent's payoffs depend in part on the indirect connections gained through her direct connections, her actions will be a response to her \emph{beliefs} of the structure of those connections. 

De Mart{\'{i}} and Zenou (2015) \cite{DeMarti2015} assume that there is a shared state of the world which determines the return individuals receive from expending private effort to interact with each other. Since this state is further assumed to be unknown, players use signals to partially infer the true state. 

This type of model arguably forms the main stream within this literature and generally focuses on network \emph{games}. That is, the objective is to understand better how games are played when the players are already connected in a network in some fashion and, as such, these models differ significantly from network \emph{formation} models.

The work of Song and van der Schaar (2015) \cite{Song2015} is a notable exception that parallels the approach of this paper. They consider not only a network formation model but also a heterogeneous agent setting in which agents have incomplete information about each others' types. Their model, however, assumes that heterogeneity is expressed in benefits and that beliefs are shared across agents. 

Galeotti, Goyal and Kamphorst (2006) \cite{Galeotti2006} note that heterogeneity in cost shapes both the level of connectedness as well as the general structure of a network, whereas heterogeneity in benefits appears to have much less of an influence on the latter.

\section{The Incomplete Information Model}

\subsection{Design}

In order to allow for homophily to be the result rather than the entry point of a model, we introduce the idea that agents belong to distinct \emph{social groups} which share observable characteristics but also have a hidden \emph{type} which can be revealed by forming a link. 


\paragraph{Agents}
Let every agent $i \in N = \{1,...,n\}$ belong to one of $K$ social groups so that $s_i = k \in K$ denotes $i$'s membership of group $k$. We define $N^k \subset N$ to be the set of all agents from group $k$. 

Let each agent also have a hidden type $t_i \in T$, where $T$ is the set of all possible types. Agents are assumed to \emph{memorise} the types of all agents whose type they learn by forming a link. Let the $n \times n$ matrix $M$ contain the memory of all agents, where $M_{ij} = 1$ if agent $i$ knows the type of agent $j$ and $M_{ij} = 0$ otherwise. 

\paragraph{Networks}

To encompass the information present within the network, we represent networks as $(N, M, g)$, where $g$ is an undirected graph. 

\paragraph{Dynamic Network Formation}

We follow Jackson and Watts (2002) \cite{Jackson2002} and explicitly define network formation as a stochastic dynamic process in which at every time step a random pair $\{i,j\}$ is selected with uniform probability. This pair is evaluated for pairwise stability and, should their relationship be unstable, a link is then either added or removed.


\paragraph{Actual Utility}

Agents are assumed to receive utility from their location within the network as follows:

\begin{equation}
    u_i(g) = \sum_{j \neq i} \delta ^{d_{ij}} - \sum_{j:\ ij\ \in\ g}c_{ij}
\label{realised_utility}
\end{equation}

where we assume that benefits decay with the geodesic distance, $d_{ij}$, given the decay factor, $0 < \delta < 1$. Further, we assume the cost of a direct connection, $c_{ij}$, depends on the agents' types, specifically:

\begin{equation}
    c_{ij} = c(t_i, t_j)
\label{general_cost}
\end{equation}

\paragraph{Expected Utility}

For prospective links, agents are required to consider \emph{expected utility} when contemplating a link with someone they do not yet know. 

This paper contributes a \emph{specific} formulation for the expected utility. First note that since agents have met everyone they are presently in a direct connection with, the uncertain component is expressed only when the incremental change from a link \emph{addition} is considered. The utility an agent expects to enjoy if the link is added can then be expressed as: 

\begin{equation}
    \begin{split}
    E_i[u_i(g+ij)] & = [\delta - \sum_{t \in T} \pi_i(t|s_j)\ c(t_i, t)] \\
                    & + [\sum_{k \notin \{i,j\}} \delta ^{d_{ik}} - \sum_{k:\ ik\ \in\ g \setminus ij}c_{ik}] 
    \end{split}
    \label{expected_gain}
\end{equation}

The first bracketed term on the right-hand side corresponds to the benefit of becoming neighbours with $j$ less the expected cost of doing so. The expected cost is an average of the costs to connect with each type $t \in T$ weighted by agent $i$'s belief -- $\pi_i(t|s_j)$ -- that agent $j$ is of type $t$ given their social group. The second bracketed term describes the net utility from the remainder of the network defined by $g+ij$.

Note that we can thus decompose the above expression into (i) the expected incremental change in utility from adding a link to agent $j$, $E_i[u_{ij}]$, and (ii) the utility she starts off with, giving the following expression:  

\begin{equation}
    E_i[u_i(g+ij)] = E_i[u_{ij}] + u_i(g)
    \label{Eu_ij}
\end{equation}

\paragraph{Beliefs}

To complete our definition of expected utilities, agent $i$'s belief that agent $j$ is of type $t \in T$ given $j$'s group membership, $s_j$, is defined as follows:

\begin{equation}
    \pi_i(t|s_j) \in \begin{cases}
        [0,1] &\text{if $i$ and $j$ do not know each other} \\    
        \{0,1\} &\text{otherwise}
    \end{cases}
    \label{beliefs}
\end{equation}

This definition captures how beliefs are uncertain if agents are unacquainted but update to reflect certainty once they meet and reveal each other's types. 


\paragraph{Stability}

We can now state a refinement of the pairwise stability conditions of Jackson and Wolinsky (1996) \cite{Jackson1996} which reflects that agents do not know each other's types \textit{ex-ante}:

\begin{equation}
    \begin{split}
        \mbox{(i)\ \ \ } & \forall ij \in g,\ u_i(g) \geq u_i(g-ij) \mbox{\ and\ } u_j(g) \geq u_j(g-ij), \mbox{\ and} \\
        \mbox{(ii)\ \ \ } & \forall ij \notin g, \mbox{\ if\ } E_i[u_{ij}] \geq 0 \mbox{\ then\ } E_j[u_{ji}] < 0
    \end{split}
    \label{expstable}
\end{equation}

Note that we assume that links only require non-negative instead of strictly positive utilities to cover cases where both agents are indifferent and which are otherwise indeterminate.

\paragraph{Further Assumptions}

Having set up the general model, it is useful to follow the cost assumption of the insider-outsider model of Galeotti et al. (2006) \cite{Galeotti2006} and simplify costs to: 

\begin{equation}
    c(t_i, t_j) = \begin{cases}
        c_L &\text{if $t_i = t_j$} \\    
        c_H &\text{otherwise}
    \end{cases}
    \label{costs}
\end{equation}

where $0 < c_L < c_H$. This allows us to simplify expected costs to:

\begin{equation}
    \begin{split}
        \sum_{t \in T} \pi_i(t|s_j)\ c(t_i, t) & = \pi_i(t_i|s_j)\ c_L + (1 - \pi_i(t_i|s_j))\ c_H \\
                                            & = \pi_i(s_j)\ c_L + (1 - \pi_i(s_j))\ c_H
    \end{split}
    \label{expcosts}
\end{equation}

where $\pi_i(s_j)$ is shorthand for $\pi_i(t_i|s_j)$.

This simplification implies that the cost of forming inter-type links only depends on the fact that types differ rather than by how much. This allows us to hone in on the relationship between beliefs and network outcomes.  

\paragraph{Complete Information}

One observation to make is that our model nests the complete information case. In particular, if agents are assumed to know each other, then their beliefs are updated to $\pi_i(s_j) \in \{0,1\}$ for all $i,j \in N$ prior to the start of the game. Expected utilities then collapse to actual utilities. Further, our stability conditions will effectively coincide with those of other complete information models, such as Jackson and Wolinsky (1996) \cite{Jackson1996} or Jackson and Rogers (2005) \cite{JacksonRogers2005}, save for our inclusion of an explicit definition for previously indeterminate links.

We can further observe that the model of Jackson and Rogers (2005) \cite{JacksonRogers2005} (with their truncation parameter set to infinity) is a special case of our model with complete information where hidden types and social groups coincide.

\subsection{The Role of Beliefs}

Our first set of propositions revolves around the role of beliefs in the decision of when to make connections, particularly for low intra-type costs. Note that the relationship between beliefs and networks is, in general, quite complex when costs are entirely unrestricted. Given our objective of studying social networks, it seems reasonable, however, to assume that like-minded individuals would want to meet. We thus focus our attention on the low intra-type cost case.

These propositions demonstrate that in many cases, beliefs are sufficient for links to initially form or not form, irrespective of gains in actual utility. This further has implications on the information agents will acquire over each other's types. 

Note that all proofs are given in Appendix \ref{appx:proof}.

\begin{proposition}
    Let $c_L \leq \delta - \delta^2$. Any two agents $i$ and $j$ must know each other in a pairwise stable equilibrium, if
    
    \begin{enumerate}[(i)]
        \item $min\{\pi_i(s_j), \pi_j(s_i)\} \geq \tfrac{c_H - (\delta - \delta^2)}{c_H - c_L}$, provided $c_H > \delta - \delta^2$, or 
        \item for any $\pi_i(s_j), \pi_j(s_i) \in [0,1]$, provided $c_H \leq \delta - \delta^2$.
    \end{enumerate}
    \label{prop1}
\end{proposition}

This proposition has an intuitive interpretation. If inter-type costs are low, then agents will always want to meet regardless of their beliefs. If inter-type costs are high, then both agents need to be sufficiently optimistic about compatibility before agreeing to meet. 

\begin{corollary}
    Let $c_L \leq \delta - \delta^2 < c_H$. There exist $\pi_i(s_j), \pi_j(s_i) \in [0,1]$ so that agents $i$ and $j$ must know each other's types in a pairwise stable equilibrium.
    \label{cor1}
\end{corollary}

This corollary is useful, as it implies that for inter-type costs that are arbitrarily high, there are beliefs that are able to persuade the agents to still meet. 

\begin{proposition}
    Consider the empty network with agents that have never met. This network cannot be pairwise stable if and only if
    \begin{enumerate}[(i)]
        \item for some $i, j \in N$, $min\{\pi_i(s_j), \pi_j(s_i)\} \geq \tfrac{c_H - \delta}{c_H - c_L}$ when $c_H > \delta$, or
        \item $c_H \leq \delta$.
    \end{enumerate}
    \label{prop2}
\end{proposition}

\begin{corollary}
    Consider a network that is initially empty with agents that have never met and let $c_L \leq \delta - \delta^2 < \delta < c_H$. Then a pairwise stable network must have: 
    
    \begin{enumerate}[(i)]
        \item no one knowing each other if for any $i,j \in N, \pi_i(s_j) < \tfrac{c_H - \delta}{c_H - c_L}$ \smallskip
        \item some agents $i,j$ that have discovered each other's types if \\
        $min\{\pi_i(s_j), \pi_j(s_i)\} \geq \tfrac{c_H - \delta}{c_H - c_L}$ \smallskip
        \item everyone knowing each other if for any $i,j \in N, \\
        min\{\pi_i(s_j), \pi_j(s_i)\} \geq \tfrac{c_H - (\delta - \delta^2)}{c_H - c_L}$
    \end{enumerate}
    \label{cor2}
\end{corollary}

This proposition and corollary are useful for simulations where the network may be initially instantiated as the empty network with agents that do not yet know each other. The proposition provides parameterisation that ensures agents will start revealing to each other their types, while the corollary provides some additional predictions about the extent to which discovery occurs. Note that they do not imply that the empty network cannot be pairwise stable, as agents may decide to terminate their links once they know their neighbours' types.

\begin{lemma}
    Consider an agent $i$ who belongs to a component $C_i$ and who is not connected to and does not know agent $j$ who belongs to a separate component $C_j$. Let $c_H > \delta + (|C_j| - 1)\ \delta^2$. In a pairwise stable equilibrium, agent $i$ will refuse to become acquainted with $j$ if:
    
    \begin{equation*}
        \pi_i(s_j) < \dfrac{c_H - (\delta + (|C_j| - 1)\ \delta^2)}{c_H - c_L}
    \end{equation*}    
    
    \label{prop-antisocial}
\end{lemma}

This lemma shows that our model is also able to support pairwise stable equilibria in which components coexist. Note that this nests the special case of a singleton refusing to connect to any component. The interpretation is, in addition, intuitive. An agent $i$ will refuse to branch out of his group, even to the best connected individual of another group, if he is sufficiently pessimistic about compatibility. 

\subsection{Comparison with Complete Information Networks}


Song and van der Schaar (2015) \cite{Song2015} derive a theorem that includes a useful relationship between incomplete and complete information networks, showing that the set of networks that can be generated and the set of networks that are stable under complete information are, respectively, subsets of those under incomplete information. While their model assumes heterogeneity in benefits, the proof of their theorem does not depend on this fact and can thus be applied to our model. We therefore restate a generalised excerpt of their theorem without proof. 

First, denote the set of all networks generated under complete information as $G^C(N)$ and those under incomplete information as $G^{IC}(N)$. Let the networks that are pairwise stable under complete and incomplete information respectively be denoted as $G^C_S(N)$ and $G^{IC}_S(N)$.

\begin{theorem}
    If $E_i[u_{ij}] \geq 0$ for all $i,j \in N$, then $G^C(N) \subset G^{IC}(N)$ and $G^C_S(N) \subset G^{IC}_S(N)$.
    \label{song1}
\end{theorem}

This allows us to derive the following proposition characterising the relationship between incomplete and complete information networks in dependence of beliefs. 

\begin{proposition}
    Let $c_L \leq \delta - \delta^2$ and $min\{\pi_i(s_j), \pi_j(s_i)\} \geq \tfrac{c_H - (\delta - \delta^2)}{c_H - c_L}$ for all $i,j \in N$. Then 
    
    \begin{enumerate}[(i)]
        \item $G^C(N) \subset G^{IC}(N)$ and $G^C_S(N) \subset G^{IC}_S(N)$, but 
        \item $G^{IC}_S(N) \not\subset G^{C}_S(N)$.
    \end{enumerate}
    \label{prop3}
\end{proposition}

Proposition \ref{prop3} proves that for low intra-type costs our model is both capable of generating any network that can be generated under complete information as well as supporting all pairwise stable networks. Further, it shows that the set of pairwise stable networks under complete information constitutes a strict subset. 

This is a significant result, as it implies that some links are formed and \emph{kept} under incomplete information that would otherwise never have survived if agents knew who they were about to connect with.

\subsection{Application to Segregation in Social Networks}


One of our key objections to the existing literature relate to an implicit assumption that homophily is an inherent preference over observable attributes. This assumption appears to suggest that segregative patterns in society which are most vividly based on outward traits -- such as ethnicity, gender or class -- are the socially optimal result of our supposed preference to be with people of similar such traits. 

A more nuanced argument could be made which points out that the groups in de Mart{\'{i}} and Zenou (2017) \cite{DeMarti2017} or the islands in Jackson and Rogers (2005) \cite{JacksonRogers2005} do not need to correspond to social groups defined by their outward appearance. Segregative patterns could be the result of people forming relationships based on preferences for some underlying trait which happens to correlate with the outward traits we observe. Homophily could, in that case, be interpreted as a resulting phenomenon rather than an inherent preference. This line of reasoning, however, does not fully convince, as the underlying traits are then necessarily assumed to be observable by everyone.

We have therefore proposed a model of a heterogeneous agent-based network formation model under uncertainty to relax this assumption and, instead, allow for people to look for hidden traits in each other based on expectations formed from their outward appearances. If these expectations are biased, as this paper contends, then social group homophily could be interpreted as a reflection of these biases rather than an inherent preference. 

The introduction of a distinction between hidden types and overt social groups provides an additional advantage in that the characterisation of efficient networks carries over from complete information models. The analysis of Jackson and Rogers (2005) \cite{JacksonRogers2005}, for instance, can be applied to our model by assuming that their islands are defined based on our hidden types rather than social groups.  

We demonstrate in Appendix \ref{appx:prop} that we are easily able to generate a segregative equilibrium which is not socially optimal.

\section{Simulation Setup}
To complement the static perspective taken thus far, we will now turn to the numerical investigation of the role of beliefs in dynamic network formation under incomplete information. We specifically focus on the properties of the networks after they have converged to a steady state.

In the following we discuss specifics of the design of our simulation and the key metrics that we are measuring.

\subsection{Initial Conditions}
From our theoretical results, it is clear that given the initial state of the system, different pairwise stable networks will form. 

The objective of this paper is to investigate the effect of incomplete information on the formation of networks. As such, the focus is not on the analysis of any particular architectures which may require specific initialisation of the network. We will therefore initialise the network as the empty network. 

\subsection{Dynamic Process}
Our simulation will closely follow the definition of dynamic network formation due to Jackson and Watts (2002) \cite{Jackson2002} -- as we did in our theoretical framework. 

Recall that network formation is thus described as a stochastic dynamic process in which a pair of agents, $\{i,j\}$, is randomly selected in every period. This pair is then evaluated for pairwise stability. 


Instead of ending the period when a stable pairing has been encountered, we instead randomly select another pair of agents until an unstable pair has been identified or until no pairs remain. We make this change to enable faster detection of a network's pairwise stability. 

Once the relationship between one pair of agents has been modified, we subsequently move on to the next period, in which a new pair of agents, $\{k,l\}$, is selected. The process converges when no unstable pairs remain.

\subsection{Terminating Conditions}
Given that dynamic processes may either converge to a steady state or diverge, we turn to the definition of the conditions under which our system comes to a halt. 

We follow Jackson and Watts (2002) \cite{Jackson2002} and make use of what they call \textit{``improving paths"}. 



Jackson and Watts (2002) \cite{Jackson2002} show that given a finite number of network states, any improving path must either terminate in a pairwise stable network or result in a cycle, which repeatedly visits the same set of networks. 

In order to capture the role of information, we will add to the original definition of improving paths and reflect that the state of the network not only includes a description of the relationship between all agents but also the information that each agent holds. Recall that we represent this information with an $n \times n$ matrix $M$, giving us networks represented as $(N, M, g)$. 


In the spirit of Jackson and Watts (2002) \cite{Jackson2002}, we define an improving path from a network $(M, g)$ to a network $(M', g')$ -- where we suppress the constant set of agents $N$ for ease of notation -- as a finite sequence of adjacent networks $(M_1, g_1), (M_2, g_2), ..., (M_K, g_K)$ with $(M_1, g_1) = (M, g)$ and $(M_K, g_K) = (M', g')$ so that for any $k \in \{1,...,K-1\}$:

\begin{equation}
    \begin{split}
        \mbox{(i)\ \ } & g_{k+1} = g_k - ij \mbox{\ for some\ } ij \mbox{\ s.t.\ } u_i(g_k - ij) > u_i(g_k), \mbox{\ or} \\
        \mbox{(ii)\ \ } & g_{k+1} = g_k + ij \mbox{\ for some\ } ij \mbox{\ s.t.\ } E_i[u_{ij}] \geq 0 \mbox{\ and\ } E_j[u_{ji}] \geq 0, \mbox{\ or}\\
        \mbox{(iii)\ \ } & \mathds{1}(M_{k+1}) > \mathds{1}(M_k). 
    \end{split}
\end{equation}

where the function $\mathds{1}$ returns the number of ones in matrix $M$.

Following Jackson and Watts (2002) \cite{Jackson2002}, we similarly define cycles as a set of networks $C$ so that for any of its networks, $(M,g), (M',g') \in C$, there exists an improving path from $(M,g)$ to $(M',g')$. A \emph{closed} cycle is then a cycle $C$ that does not contain any network which lies on an improving path to a network outside of $C$.

If the improving path terminates, we can thus terminate the simulation as convergence to a pairwise stable solution has been achieved. If the improving path has a closed cycle, then our simulation will never converge and we thus also terminate. 

\subsection{Belief Mechanism}
One of the key features of our network formation model is a flexible specification of agent beliefs. This is motivated by substantial evidence that people exercise heuristic thinking in many aspects of their lives \cite{Chen2007,Gant1984,Grether1980,Kahneman1973,Krawczyk2019,Tversky1974} so that beliefs may be biased and thus differ from rational expectations. While we draw inspiration from that body of literature, our theoretical framework, however, does not assume any particular belief mechanism.

We therefore use a simulation-based approach to investigate the role of biased beliefs in generating segregation. To do so, we consider a mechanism based on heuristic thinking whereby agents judge each other's compatibility based on similarity. We specifically assume that agents possess base beliefs that are rationally anchored but tend to be more optimistic when judging people of their own social group and less so for any other social group. We express this formally as follows:

\begin{equation}
    \pi_i(s_j) = \begin{cases}
        \dfrac{n_{s_jt_j}}{n_{s_j+}} + (1-\dfrac{n_{s_jt_j}}{n_{s_j+}}) \gamma &\text{if $s_i = s_j$} \\    
        \dfrac{n_{s_jt_j}}{n_{s_j+}} - \dfrac{n_{s_jt_j}}{n_{s_j+}} \gamma                            &\text{otherwise}
    \end{cases} 
\end{equation}

where $n_{s_jt_j}$ is the cardinality of the set of agents from group $s_j$ and of type $t_j$ and $n_{s_j+}$ is the cardinality of the set of agents from group $s_j$. Their ratio corresponds to the true probability that an agent belongs to type $t_j$ in group $s_j$ and therefore equals the beliefs an agent under rational expectations would hold. The bias enters via an adjustment factor $\gamma \in [0,1]$ moving beliefs upward for agents of the same group and downward otherwise. The adjustment is normalised to ensure beliefs remain in the unit interval.

For our base case, we assume $\gamma$ is drawn from a positively skewed beta distribution: 

\begin{equation}
    \gamma \sim Beta(\alpha,\beta) \ \text{with} \ \alpha = 1, \beta > 1
\end{equation}

Note that the shape parameter $\beta$ controls the degree of bias, i.e., the extent to which agent beliefs deviate from rational beliefs. In our proposed base case parameterisation we choose $\beta = 7$ which implies that c.52\% of agents do not deviate more than 10\% and c.79\% do not deviate more than 20\% from rational beliefs. We show an illustration of the proposed belief distribution in Appx. \ref{appx:figs}.

\subsection{Metrics}



\paragraph{Segregation Indices}


We make use of the generalised Freeman's segregation index as derived in Bojanowski and Corten (2014) \cite{Bojanowski2014} which is given by:

\begin{equation}
    S_{F} = 1 - \dfrac{pn(n-1)}{(\Sigma_kn_{+k})^2 - \Sigma_kn^2_{+k}}
    \label{eq:freeman}
\end{equation}

where $n$ and $n_{+k}$ denote the cardinality of the set of all agents and the set of agents of type $k$ respectively. Further, $p$ denotes the proportion of inter-group links, which is given by the following expression:

\begin{equation}
    p = \dfrac{\Sigma_{g,h;g \neq h}m_{gh1}}{m_{++1}}
    \label{eq:intergroup}
\end{equation}

where $m_{gh1} = |\{ij \in g| i\in N^g; j \in N^h\}|$ and $m_{++1} = \Sigma_g\Sigma_h m_{gh1}$ are, respectively, the number of inter-group links and the number of all links.

Given the random graph benchmark used in Freeman's segregation index, it can be interpreted as a measure of the segregation introduced by agent preferences guiding link formation. Our model of agent-based network formation under incomplete information, however, introduces an additional information effect, which would be confounded in Freeman's segregation index.

In order to disentangle information effects from agent preferences we are therefore proposing additional \emph{incremental segregation indices} defined as follows: 

\begin{equation}
    S_{IS} = \dfrac{p_{base} - p_{biased}}{p_{base}}
    \label{eq:incremental}
\end{equation}

where $p_{biased}$ denotes the proportion of inter-group links under our network with biased beliefs and where $p_{base}$ represents the proportion of such links in a baseline network. We use a network under complete information and a network under rational expectations for our baselines. The incremental segregation indices with these baselines serve as a gauge of how much of Freeman's segregation index is attributable to biased beliefs, incomplete information and agent preferences. 

To enhance comparability between the networks, we hold the random seed constant for each biased belief network and its baseline networks. 

\paragraph{Degree Centrality}

In order to measure how well-connected the graph is, we make use of the notion of \emph{degree centrality}. Since we wish to focus on the macro-characteristics of the network as a whole, we will specifically report the average of the degree centrality across all agents. 








\paragraph{Discovery}

A distinguishing feature of our model lies in the information matrix capturing varying degrees of information present in the system. As agents become less likely to form links when the expected costs of doing so increase, the amount of information they will discover will suffer as a result. We therefore report on the proportion of discovery, which is calculated as the following ratio: 
\begin{equation}
    discovery(g) = \dfrac{\Sigma_{i}\Sigma_{j}M_{ij}}{n^2} 
    \label{eq:degree-centrality}
\end{equation}

which corresponds to the sum of the number of agents each person knows, normalised by the maximum possible such sum given a population of $n$ agents.




\section{Simulation Results}

We present our results, which were generated with the parameterisation summarised in Table \ref{table:case1}. We choose a network size for which convergence can still be reasonably expected in a real life setting, considering that all pairs of agents need to have interacted a potentially large number of times.

We focus in the main body on networks with two social groups which are each equally divided between two hidden types. In Apps. \ref{appx:sim:grp} and \ref{appx:sim:typ} we compare and discuss results from compositional changes to the network using different distributions of social groups and hidden types. In general, these additional results indicate that segregation dynamics are not significantly influenced by social groups but do depend on the composition of hidden types. We, however, show that the case of two hidden types has the advantage of incentivising network formation and thus serves as a useful counterpoint to the distortive and segregative influence of biased beliefs. 

We focus on the network dynamics by varying one type of cost at a time to investigate their independent effects. In addition, we are holding the decay factor, $\delta$, constant in all cases, as results mainly depend on the relative magnitude of benefits and costs. 

All simulations were repeated 30 times. Specimen networks are illustrated in Appx. \ref{appx:figs}. Averaged results are reported below.

\begin{table}[hbt!]
\centering
\caption{Simulation parameters.}
    \begin{tabularx}{0.75\textwidth}{l Y}
    \addlinespace
    \toprule
    {Parameter}                     & {Value}                   \\ \midrule
    Number of nodes                & 48                        \\
    Social groups                   & 2 of size 24               \\ 
    Hidden types                    & 2 of size 12 per group             \\ \midrule
    Decay factor $\delta$            & 0.7                      \\ 
    Intra-type cost $c_L$ (base case)   & 0.2                   \\
    Inter-type cost $c_H$ (base case)   & 1.0                   \\ \bottomrule
    \end{tabularx}
\label{table:case1}
\end{table}

\subsection{Segregation under uncertainty}
\begin{figure}
    \centering
    \includegraphics[width=\linewidth]{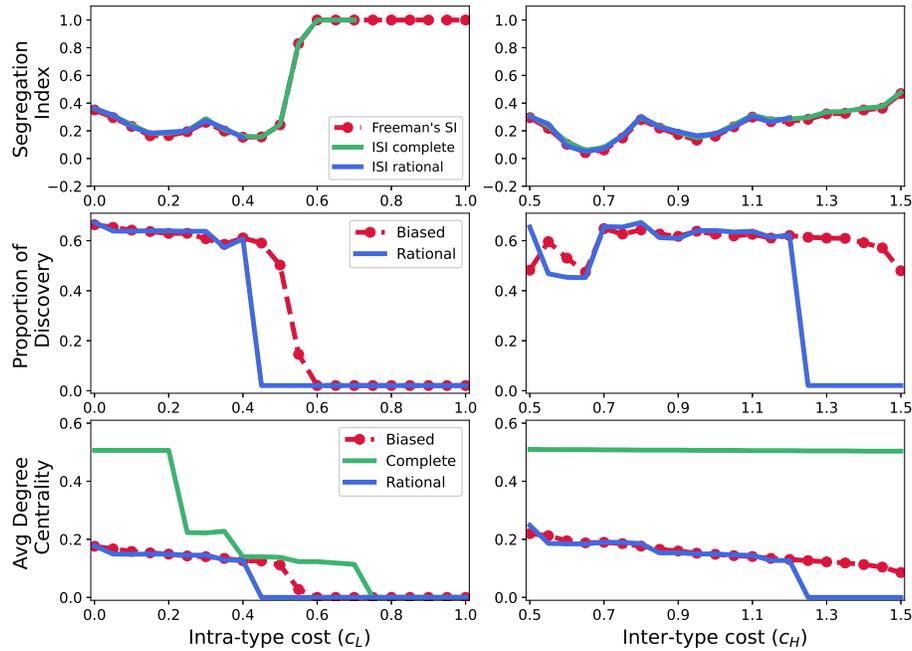}
    \caption{Biased beliefs sustain segregation as each cost is varied. Note that incremental segregation cannot be computed where the baseline network has collapsed to singletons (i.e., $p_{base} = 0$). Freeman's segregation index approaches 1 when the biased network collapses to singletons.}
\label{fig:base}
\end{figure}

In Fig. \ref{fig:base} we compare the network dynamics under biased beliefs with those under rational expectations and complete information. 

The segregation indices in the top panels confirm our hypothesis that biased beliefs play an important role in segregation under uncertainty. Freeman's segregation index remains distinctly positive as intra-type and inter-type costs change. Moreover, as both the incremental segregation indices with respect to rational expectations and complete information closely track Freeman's index, this suggests that the segregation is not a result of uncertainty alone or agent preferences respectively.

The middle panels provide a comparison of how much agents are willing to link up and reveal each other's types under biased beliefs and rational expectations. This in itself does not imply that the links that are initially formed survive once types are revealed. We observe that biased beliefs sustain connection-forming between agents for both higher intra-type and inter-type costs compared to the case under rational expectations.   

Intuitively, biased beliefs imply that agents may hold different beliefs, specifically there will exist both more and less optimistic agents in a network under beliefs following a distribution than in one under rational expectations where all agents share the same rational beliefs. The existence of more optimistic agents is significant, as they are the ones that will still be willing to risk forming a link under higher costs. As such, biased beliefs induce connections to form in higher cost ranges whereas they wouldn't under rational expectations. 

The lower panels show how well connected the network is, that is, how many connections between agents survive in equilibrium. Firstly, we can observe that networks under biased beliefs and rational expectations appear to have a similar degree of connectedness for the lower part of the cost ranges. Given the evolution of the segregation indices, this suggests that biased networks support as many links as networks with rational expectations, but with a greater number of links between agents of the same social group. 

Secondly, we can observe that the greater willingness to connect at higher costs under biased beliefs compared to under rational expectations also results in a correspondingly later collapse of the biased network. 

Lastly, networks under complete information are better connected in equilibrium. This is intuitive, as unconnected agents of the same type will always form and maintain a link as long as the intra-type cost lies below the decay factor -- a decision that is independent of inter-type costs as indicated by the flat degree centrality curve on the right-hand side of the panel. 

\vspace{-10pt}
\vspace{-10pt}

\begin{figure}
    \centering
    \includegraphics[width=\linewidth]{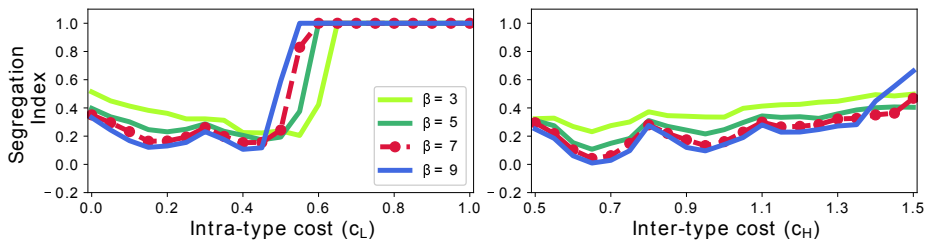}
    \caption{An increase in bias (a reduction in the $\beta$ shape parameter) leads to greater segregation. Freeman's segregation index approaches 1 when the network collapses to singletons.}
\label{fig:bflex-redux}
\end{figure}

In Fig. \ref{fig:bflex-redux} we vary the degree of bias by changing the $\beta$ shape parameter of the beta distribution seeding our beliefs where a lower $\beta$-value corresponds to a greater degree of bias. We observe that while increasing the bias is clearly associated with a step-up in the segregation index, a reduction in the bias still yields distinctly segregative results. A full comparison of network dynamics is shown in Appx. \ref{appx:sim:bias}.

\vspace{-10pt}
\vspace{-10pt}




\section{Conclusion}

We have contributed an agent-based network formation model for the study of group segregation relaxing the common assumption of complete information. We have done so by introducing notions of expectations and beliefs that are allowed to be heterogeneous across agents, drawing inspiration from the field of psychology. We have shown that our model is rich enough to not only generate those networks that a complete information equivalent could support but indeed is capable of generating an even more diverse set of networks. Beyond this, we have demonstrated that segregative equilibria can be supported in our model when agents are sufficiently biased. We have proceeded to a simulation-based approach to investigate the dynamics of segregation and found clear evidence pointing to the role of biased beliefs in generating segregation under incomplete information. 

This has not been an exhaustive study of beliefs. The belief mechanisms we have assumed throughout have been static to enable our focus on the presence and absence of complete information. A promising avenue of further research would be to imbue our mechanisms with greater realism. A first step could be to assume that beliefs gradually update as agents meet each other. Similarly, agents may not have perfect memory and forget each other if enough time has passed. A second line of extensions could emanate from incorporating more complex relationships between social groups and beliefs, where a spatial cost topology, such as in Johnson and Gilles \cite{Johnson2000}, could be applied to expected costs. This could be useful in moving the focus from overall segregative patterns as in this paper to studying segregation between specific groups. A third topic of future research could revolve around the stability conditions underlying our model. While pairwise stability offers some attractive properties, it would be worthwhile investigating the effect of coalitional stability concepts such as that proposed by Jackson and van den Nouweland (2005) \cite{JacksonNouweland2005}.

%
%
%
\bibliographystyle{splncs04}
\bibliography{references}

\begin{thebibliography}{10}
\providecommand{\url}[1]{\texttt{#1}}
\providecommand{\urlprefix}{URL }
\providecommand{\doi}[1]{https://doi.org/#1}

\bibitem{DBLP:conf/atal/AitsCT19}
Aits, D., Carver, A., Turrini, P.: Group segregation in social networks. In:
  Proceedings of the 18th International Conference on Autonomous Agents and
  MultiAgent Systems, {AAMAS} '19, Montreal, QC, Canada, May 13-17, 2019. pp.
  1524--1532. International Foundation for Autonomous Agents and Multiagent
  Systems (2019)

\bibitem{Blumenstock2019}
Blumenstock, J.E., Chi, G., Tan, X.: {Migration and the value of social
  networks}. CEPR Discussion Paper p. No. DP13611 (2019)

\bibitem{Bojanowski2014}
Bojanowski, M., Corten, R.: {Measuring segregation in social networks}. Social
  Networks  \textbf{39}(1),  14--32 (2014)

\bibitem{Centola2007}
Centola, D., Gonzlez-Avella, J.C., Egu{\'{i}}luz, V.M., {San Miguel}, M.:
  {Homophily, cultural drift, and the co-evolution of cultural groups}. Journal
  of Conflict Resolution  \textbf{51}(6),  905--929 (2007)

\bibitem{Chay2013}
Chay, K., Munshi, K.: {Black Networks After Emancipation: Evidence from
  Reconstruction and the Great Migration}. Working Paper  (2013)

\bibitem{Chen2007}
Chen, G., Kim, K.A., Nofsinger, J.R., Rui, O.M.: {Degrees of Uncertainty: An
  Overview and Framework for Future Research on Experience-Based Choice}.
  Journal of Behavioral Decision Making (20),  425--451 (2007)

\bibitem{Galeotti2006}
Galeotti, A., Goyal, S., Kamphorst, J.: {Network formation with heterogeneous
  players}. Games and Economic Behavior  \textbf{54}(2),  353--372 (2006)

\bibitem{Gant1984}
Gant, M.M., Davis, D.F.: {Mental Economy and Voter Rationality: The Informed
  Citizen Problem in Voting Research}. The Journal of Politics  \textbf{46}(1),
   132--153 (1984)

\bibitem{Granovetter1983}
Granovetter, M.: {The Strength of Weak Ties: A Network Theory Revisited}.
  Sociological Theory  \textbf{1},  201--233 (1983)

\bibitem{Grether1980}
Grether, D.M.: {Bayes Rule as a Descriptive Model: The Representativeness
  Heuristic}. The Quarterly Journal of Economics  \textbf{95}(3), ~537 (1980)

\bibitem{JacksonNouweland2005}
Jackson, M.O., van~den Nouweland, A.: {Strongly stable networks}. Games and
  Economic Behavior  \textbf{51}(2 SPEC. ISS.),  420--444 (2005)

\bibitem{JacksonRogers2005}
Jackson, M.O., Rogers, B.W.: {The Economics of Small Worlds}. The Economic
  Journal  \textbf{3}(2-3),  617--627 (2005)

\bibitem{Jackson2002}
Jackson, M.O., Watts, A.: {The evolution of social and economic networks}.
  Journal of Economic Theory  \textbf{106}(2),  265--295 (2002)

\bibitem{Jackson1996}
Jackson, M.O., Wolinsky, A.: {A strategic model of social and economic
  networks}. Journal of Economic Theory  \textbf{71}(1),  44--74 (1996)

\bibitem{JacksonXing2014}
Jackson, M.O., Xing, Y.: {Culture-dependent strategies in coordination games}.
  Proceedings of the National Academy of Sciences of the United States of
  America  \textbf{111}(SUPPL.3),  10889--10896 (2014)

\bibitem{JacksonYariv2007}
Jackson, M.O., Yariv, L.: {Diffusion of behavior and equilibrium properties in
  network games}. American Economic Review  \textbf{97}(2),  92--98 (2007)

\bibitem{Johnson2000}
Johnson, C., Gilles, R.P.: {Spatial social networks}. Review of Economic Design
   \textbf{5}(3),  273--299 (2000)

\bibitem{Kahneman1973}
Kahneman, D., Tversky, A.: {On the psychology of prediction}. Psychological
  Review  \textbf{80}(4),  237--251 (1973)

\bibitem{Krawczyk2019}
Krawczyk, M.W., Rachubik, J.: {The representativeness heuristic and the choice
  of lottery tickets: A field experiment}. Judgment and Decision Making
  \textbf{14}(1),  51--57 (2019)

\bibitem{DeMarti2015}
de~Mart{\'{i}}, J., Zenou, Y.: {Network games with incomplete information}.
  Journal of Mathematical Economics  \textbf{61},  221--240 (2015)

\bibitem{DeMarti2017}
de~Mart{\'{i}}, J., Zenou, Y.: {Segregation in Friendship Networks}.
  Scandinavian Journal of Economics  \textbf{119}(3),  656--708 (2017)

\bibitem{McPherson2001}
McPherson, M., Smith-Lovin, L., Cook, J.M.: {Birds of a Feather: Homophily in
  Social Networks}. Annual Review of Sociology  \textbf{27},  415--444 (2001)

\bibitem{Munshi2011}
Munshi, K.: {Strength in numbers: Networks as a solution to occupational
  traps}. Review of Economic Studies  \textbf{78}(3),  1069--1101 (2011)

\bibitem{Munshi2016}
Munshi, K., Rosenzweig, M.: {Networks and Misallocation: Insurance, Migration,
  and the Rural-Urban Wage Gap}. American Economic Review  \textbf{106}(1),
  46--98 (2016)

\bibitem{Schelling1969}
Schelling, T.C.: {Models of Segregation}. The American Economic Review
  \textbf{59}(2),  488--493 (1969)

\bibitem{Schelling1971}
Schelling, T.C.: {Dynamic Models of Segregation}. Journal of Mathematical
  Sociology  \textbf{1},  143--186 (1971)

\bibitem{Skopek2011}
Skopek, J., Schulz, F., Blossfeld, H.P.: {Who contacts whom? Educational
  homophily in online mate selection}. European Sociological Review
  \textbf{27}(2),  180--195 (2011)

\bibitem{Song2015}
Song, Y., van~der Schaar, M.: {Dynamic network formation with incomplete
  information}. Economic Theory  \textbf{59}(2),  301--331 (2015)

\bibitem{Tversky1974}
Tversky, A., Kahneman, D.: {Judgment under Uncertainty : Heuristics and
  Biases}. Science  \textbf{185},  1124--1131 (1974)

\bibitem{Verdier2017}
Verdier, T., Zenou, Y.: {The role of social networks in cultural assimilation}.
  Journal of Urban Economics  \textbf{97},  15--39 (2017)

\bibitem{Zenou2015}
Zenou, Y.: {A dynamic model of weak and strong ties in the labor market}.
  Journal of Labor Economics  \textbf{33}(4),  891--932 (2015)

\end{thebibliography}

\clearpage
\appendix
\section{Non-optimal Segregative Equilibrium}
\label{appx:prop}

In this appendix, we show that our model supports segregative equilibria that are not socially optimal. 

\begin{proposition}
    Let all agents be of the same hidden type but belong to K separate social groups, $N^{1}, N^{2}, ..., N^{K} \subset N$ and let $c_L \leq \delta - \delta^2$.
    
    Assume that beliefs are such that:
    \begin{enumerate}[(i)]
        \item for $i,j \in N^{k},\ \pi_i(s_j) \geq \tfrac{c_H - (\delta - \delta^2)}{c_H - c_L}$
        \item for $i \in N^{k}, j \in N^{s}, k \neq s,\ \pi_i(s_j) < \tfrac{c_H - (\delta + (n_s - 1)\ \delta^2)}{c_H - c_L}$
    \end{enumerate}
    
    Starting a new network from the empty network, agents will be perfectly segregated in a pairwise stable equilibrium that is not socially optimal.
    
    \label{prop-suboptimal}
\end{proposition}

\begin{proof}
    Start with the empty network and take any agent $i$ belonging to some social group $k$. 
    
    Given that $i$'s beliefs comply with the first condition for any $j \in N^{k}$, Prop. \ref{prop1} implies that agent $i$ will meet all such agents $j$, that is, he will want to meet everyone from his social group. Since all agents are of the same hidden type and $c_L \leq \delta - \delta^2$, all such meetings result in lasting connections. 
    
    Given that $i$'s beliefs comply with the second condition for any $j \in N^{s}, k \neq s$, Lemma \ref{prop-antisocial} implies that $i$ will refuse to meet $j$ no matter how well-connected she is within her group.
    
    Given $c_L \leq \delta - \delta^2$, the efficient network must be fully connected amongst agents of the same type, as the incremental utility of forming a link will be at least $u_{ij} \geq \delta - \delta^2 - c_L \geq 0$. Yet we have just shown that agents stay within their social groups, giving the proposition.
\end{proof}
\clearpage
\section{Proofs}
\label{appx:proof}

This appendix contains the proofs for the statements in the main body of the paper.

\subsection{Proof of Proposition \ref{prop1}}
\begin{proof}
    To prove this by contradiction, suppose there is a pairwise stable equilibrium in which agents $i$ and $j$ have not met. 
    
    W.l.o.g, the least amount of utility agent $i$ expects to gain from meeting $j$ corresponds to the case when $i$ and $j$ are already as closely connected as possible without being neighbours, i.e., $d_{ij} = 2$. The expected gain is therefore non-negative if: 
    \begin{equation*}
        \delta - [\pi_i(s_j)\ c_L + (1 - \pi_i(s_j))\ c_H] - \delta^2 \geq 0
    \end{equation*}
    
    where the first two terms are the benefit of becoming neighbours less the expected cost of doing so, and where the last term is the lost benefit of being closely but indirectly connected.  

    This expression can be rewritten as:
    \begin{equation*}
        \pi_i(s_j) \geq \dfrac{c_H - (\delta - \delta^2)}{c_H - c_L}
    \end{equation*}
    
    which gives us a contradiction, as agents $i$ and $j$ must therefore have met. Statements 1 and 2 then follow directly.
\end{proof}

\subsection{Proof of Corollary \ref{cor1}}
\begin{proof}
    This follows from Prop. \ref{prop1}, as $\tfrac{c_H - (\delta - \delta^2)}{c_H - c_L} \leq 1$.
\end{proof}

\subsection{Proof of Proposition \ref{prop2}}
\begin{proof}
    To prove this by contradiction, suppose the empty network with entirely ignorant agents was pairwise stable. Take any agents $i$ and $j$, who must be unacquainted with one another, so that the expected utility of forming a link for agent $i$ is therefore always non-negative if:
    
    \begin{equation*}
        \delta - \pi_i(s_j)\ c_L - (1 - \pi_i(s_j))\ c_H \geq 0
    \end{equation*}
    
    This is equivalent to:

    \begin{equation*}
        \pi_i(s_j) \geq \dfrac{c_H - \delta}{c_H - c_L}
    \end{equation*}
    
    This results in a contradiction and a link will then be formed if agent $j$'s expected utility is also non-negative. The proposition thus follows by symmetry. 
\end{proof}

\subsection{Proof of Corollary \ref{cor2}}
\begin{proof}
    This corollary follows from Prop. \ref{prop1} and Prop. \ref{prop2}.
\end{proof}

\subsection{Proof of Lemma \ref{prop-antisocial}}
\begin{proof}
    Suppose agent $i$ contemplates a link with agent $j$. The highest net utility agent $i$ can receive from becoming connected to $j$ occurs when $j$ is himself directly connected to all individuals in his component $C_j$, so that $i$ could become indirectly connected to everyone in $C_j$. The expected incremental gain of connecting is therefore negative if:
    
    \begin{equation*}
        \delta + (|C_j| - 1)\ \delta^2 - \pi_i(s_j)\ c_L - (1 - \pi_i(s_j))\ c_H < 0
    \end{equation*}
    
    which shows the lemma. 

\end{proof}

\subsection{Proof of Proposition \ref{prop3}}
\begin{proof}
    Statement 1 -- Note that Prop. \ref{prop1} implies that it must be the case that $E_i[u_{ij}] \geq 0$ for all $i,j \in N$. Thm. \ref{song1} thus gives us the first statement.

    Statement 2 -- We show this by contradiction. Suppose it was the case that $G^{IC}_S(N) \subset G^{C}_S(N)$.
    
    Lemma \ref{prop-antisocial} shows that under incomplete information some agents would stay singletons for inter-type costs that are high enough and pessimistic enough beliefs. Let this network be denoted $g^o \in G^{IC}_S(N)$.
    
    Under complete information, agents would always connect with anyone from their own type, as the least amount of utility they would receive is:
    
    \begin{equation*}
        u_{ij} \geq \delta - \delta^2 - c_L \geq 0
    \end{equation*}
    
    Singletons, therefore, do not survive under complete information, that is, $g^o \notin G^{C}_S(N)$. Hence $G^{IC}_S(N) \not\subset G^{C}_S(N)$. 
\end{proof}
\clearpage
\section{Additional Simulations}
We discuss a number of ancillary simulations in this appendix. 

\subsection{Varying the Degree of Bias}
\label{appx:sim:bias}

\begin{figure}[htb!]
    \centering
    \includegraphics[width=\linewidth]{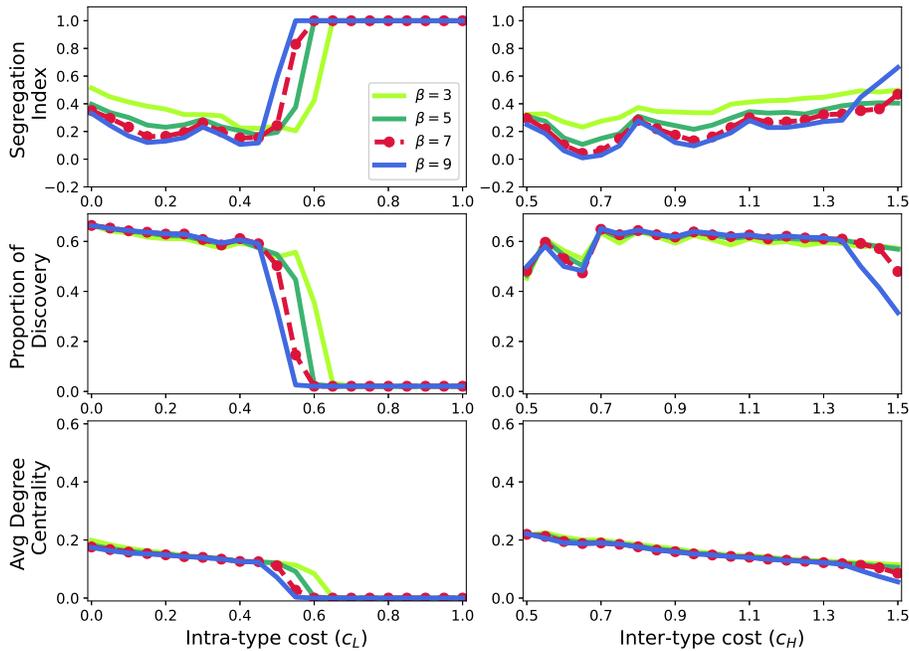}
    \caption{Increasing bias results in greater segregation. Freeman's segregation index approaches 1 when the biased network collapses to singletons.}
\label{fig:bflex}
\end{figure}

In Fig. \ref{fig:bflex} we show a full comparison of network dynamics under different levels of bias induced by changes to the $\beta$ shape parameter. We can observe that the networks with more biased agents reflect a step-up in segregation. Interestingly, such networks tend to collapse at a slightly later point when intra-type costs are increased. Intuitively, more biased agents are more willing to form connections with peers from the same social group, as they are more optimistic that such agents would also have the same hidden type. 

\subsection{Compositional Changes: Social Groups}
\label{appx:sim:grp}

In this section, we compare network dynamics under different social groupings of the network. An overview of the network compositions is shown in Table \ref{table:grp}. 

\begin{table}[hbt!]
\centering
\caption{Network compositions. Entries of arrays indicate size of a particular social group.}
    \begin{tabularx}{\textwidth}{l Y Y Y}
    \addlinespace
    \toprule
    {Parameter} & {Base} & {4 Groups} & {Imbalanced} \\ \midrule
    Social group size & [24, 24] & [12, 12, 12, 12] & [12, 36] \\ 
    Hidden type distribution & \multicolumn{3}{c}{Even split within each group}  \\           \bottomrule
    \end{tabularx}
\label{table:grp}
\end{table}

\begin{figure}[htb!]
    \centering
    \includegraphics[width=\linewidth]{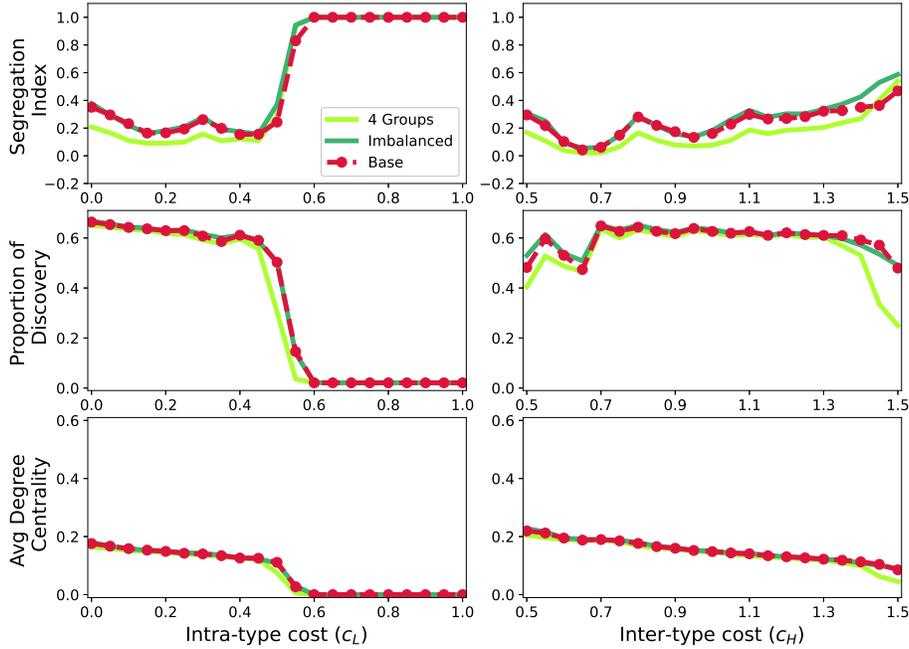}
    \caption{Segregative network dynamics under different social groupings of the network. Freeman's segregation index approaches 1 when the network collapses to singletons.}
\label{fig:grp}
\end{figure}

In Fig. \ref{fig:grp} we can observe that changes to the social groups do not seem to have a significant impact on segregative dynamics, as they closely align to our base case results.

Note that Freeman's segregation index is calculated with respect to the number of inter-group links expected in a randomly formed network. An increase in the number of social groups or a re-balancing to create majority and minority groups result in adjustments to the expected number of random inter-group links. The absolute number of inter-group link will therefore have changed between the different compositions of the network, yet the relative proportion to the respective random networks must have largely remained constant. 

\subsection{Compositional Changes: Hidden Types}
\label{appx:sim:typ}

In this section, we compare network dynamics under different distributions of hidden types. An overview of the network compositions is shown in Table \ref{table:types}. 

\begin{table}[hbt!]
\centering
\caption{Network compositions. Array entries in the first row indicate size of a particular social group. Array entries in the second row indicate size of a particular hidden type \emph{given} a social group.}
    \begin{tabularx}{\textwidth}{l Y Y Y Y}
    \addlinespace
    \toprule
    {Parameter} & {Base} & {4 Types} & {Imbalanced} & {Correlated} \\ \midrule
    Social group size & \multicolumn{4}{c}{[24, 24]}\\ 
    \makecell[l]{Hidden type\\ distribution} & \makecell{[12, 12] \\ per group} & \makecell{[6, 6, 6, 6] \\ per group} & \makecell{[18, 6]\\ per group} & \makecell{Group 1: [18, 6] \\ Group 2: [6, 18]} \\           \bottomrule
    \end{tabularx}
\label{table:types}
\end{table}

\begin{figure}[htb!]
    \centering
    \includegraphics[width=\linewidth]{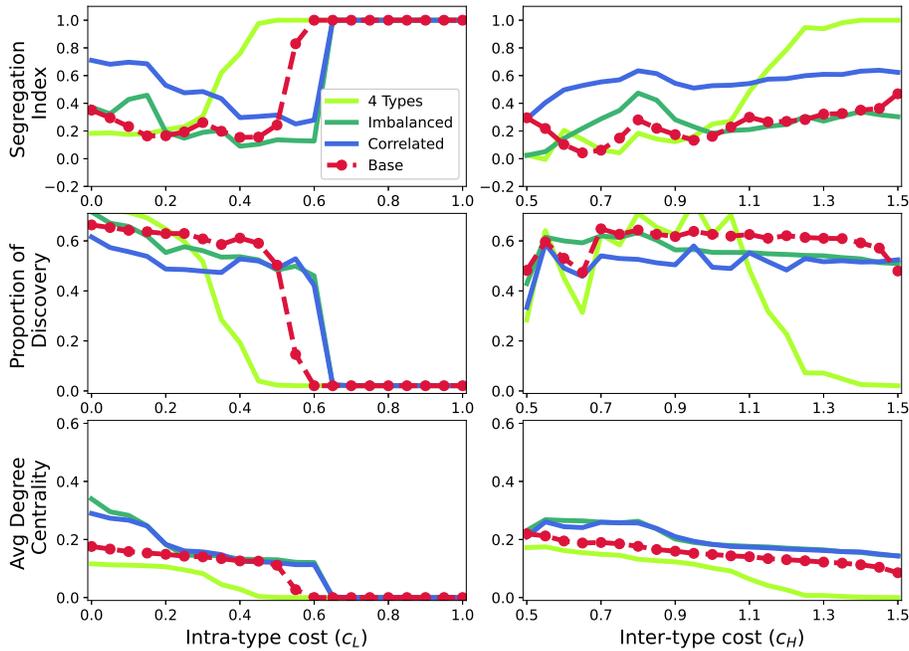}
    \caption{Segregative network dynamics under different hidden type distributions. Freeman's segregation index approaches 1 when the network collapses to singletons.}
\label{fig:type}
\end{figure}

In Fig. \ref{fig:type} we observe that network dynamics are sensitive to the distribution of hidden types. This is expected given their direct relationship to agent preferences.

The results indicate that increasing the number of types appears to have a disincentivising effect on network formation. Intuitively, a larger number of rival types dilutes incentives to form connections, as all connections with agents of a different type are penalised. Networks with 4 hidden types thus show a rapid rise in the segregation index as costs increase and collapse to singletons comparatively early.

Initialising the network with a type that represents the majority in each social group (the imbalanced case) appears to not impact the dynamics significantly. We can observe that these networks tend to be slightly better connected and collapse to singletons at slightly higher intra-type costs compared to the base case.  

In contrast, in the correlated case in which each social group is dominated by a different hidden type, the corresponding results reflect a marked increase in segregation at a comparable level of connectedness. This is not surprising, as this case naturally encourages segregation given that agents of the majority type in a given social group would \emph{rationally} prefer connections to the same social group, all things being equal. 
\clearpage
\section{Additional Figures}
\label{appx:figs}

\begin{figure}[htb!]
    \centering
    \includegraphics[width=0.6\linewidth]{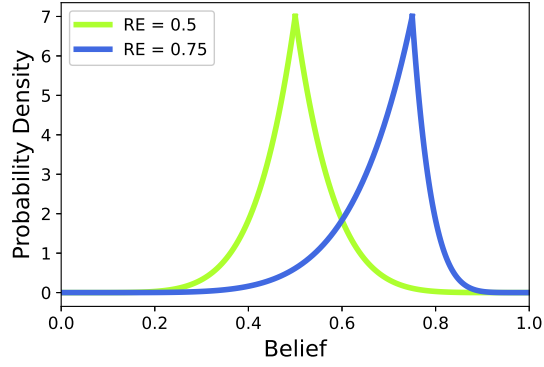}
    \caption{Probability density functions underlying the biased belief mechanism anchored by different rational expectations with the adjustment factor $\gamma \sim Beta(1,7)$. Beliefs are drawn from the right-hand side of the peak for agents of the same social group and from the left-hand side of the peak otherwise.}
\label{fig:beta}
\end{figure}

\begin{figure}[htb!]
    \centering
    \includegraphics[width=\linewidth]{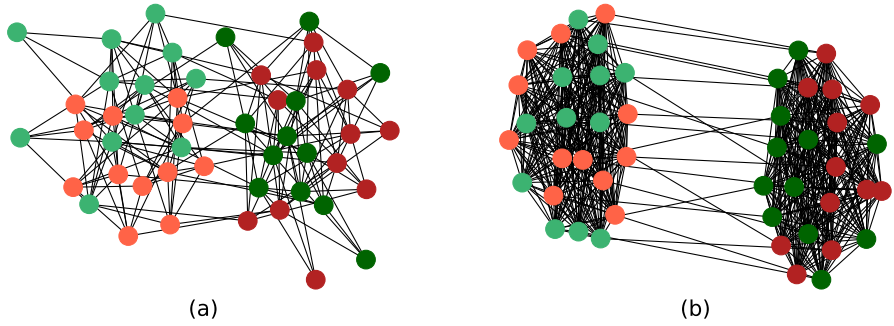}
    \caption{Example networks generated with the baseline parameterisation in Table \ref{table:case1} using (a) biased beliefs and (b) complete information. Colours (green vs red) indicate social groups, tones (light vs dark) denote hidden types.}
\label{fig:examples}
\end{figure}

\end{document}